\documentclass[aps,pra,twocolumn,showpacs,showkeys,superscriptaddress,reprint,floatfix]{revtex4-1} 
\usepackage{graphics} 
\usepackage{graphicx} 
\usepackage{amsmath} 
\usepackage{amssymb}
\usepackage{color}
\usepackage[caption=false]{subfig}

\begin{document}

\title{Mass imbalance in the ionic Hubbard model: a DRMG study}
\author{D. C. Padilla-Gonz\'alez}%
\author{R. Franco}%
\author{J. Silva-Valencia}%
\email{jsilvav@unal.edu.co}
\affiliation{Departamento de F\'{\i}sica, Universidad Nacional de Colombia, A. A. 5997 Bogot\'a, Colombia.}

\date{\today}

\begin{abstract}
We investigated the ionic Hubbard model with mass imbalance in  one dimension, using the density matrix renormalization group method. This model exhibits a band insulator phase and an antiferromagnetic one, both with a finite spin gap. We found that this quantum phase transition is continuous, unalike the previous mean-field theory result. The von Neumann block entropy is maximum at the critical points, a fact that we used to build the phase diagram.
\end{abstract}


\maketitle

%
\section{\label{sec1}Introduction}
The area of ultracold atoms has become a successful laboratory for many ideas and concepts in physics, not only allowing testing but also extending them~\cite{IBloch-RMP08,Esslinger-AR10,IBloch-NP12,Gross-S17}. A little over a decade ago, for the first time  it was possible to confine fermionic atoms in optical lattices, achieving degenerate configurations where the Mott insulator state predicted by the Hubbard model was observed~\cite{Jordens-N08}. This fact led to observing the superfluid-Mott insulator transition~\cite{Greif-Science16}, SU($N>2$) physics~\cite{Taie-NP12}, and antiferromagnetic correlations in two dimensions~\cite{Hart-N15}, among other phenomena.\par 
The achievement of having simultaneously confined different types of atoms opened the possibility of studying new phenomena~\cite{Taglieber-PRL08,Spiegelhalder-PRL09,Tiecke-PRL10,Taie-PRL10,Trenkwalder-PRL11,Kohstall-N12,Jotzu-PRL15}, for instance those generated by the difference in the masses of atoms (mass imbalance), which is relevant in fermionic superfluity~\cite{Orso-PRA08}, atom-dimer attraction~\cite{Jag-PRL14}, and the Fulde, Ferrell, Larkin, and Ovchinnikov (FFLO) mechanism that enables superconductivity~\cite{Fulde64,Larkin65}.\par 
Other model emulated in ultracold atom setups is the ionic Hubbard model (IHM), which extend the Hubbard one by adding an energy offset between next-neighbor sites. The ionic Hubbard model has been widely studied~\cite{Fabrizio-PRL99,Torio-PRB01,YZZhang-PRB03,Kampf-PRB03,Manmana-PRB04,Aligia-PRB05,Otsuka-PRB05,Refolio-JPCM05,Byczuk-PRB09,AGo-PRB11,DiLiberto-PRB14,AJKim-PRB14,Bag-PRB15,Murcia-Correa-JPCS16,Loida-PRL17,Chattopadhyay-PRB19}, and recently two-dimensional realizations of it (with square and honeycomb geometries) have been achieved~\cite{DiLiberto-NC14,Messer-PRL15}.\par 
The interplay of local interactions and the mass-imbalance in different lattice geometries, which lead to new physics, have been examined in ultracold experiments~\cite{Oppong-Arxiv20} and through diverse theoretical approaches~\cite{Cazalilla-PRL05,Garg-PRL06,Winograd-PRB2011,Dao-PRB12,YHLiu-PRB15,Philippa-EPJB17,Shahbazy-CJP19,Rammelmuller-SP20,YHe-Arxiv21}. Recently, a new ingredient was added to this competition: the band energy offset, i.e. an ionic Hubbard model was considered with mass imbalance using mean-field theories~\cite{Sekania-PRB17,DALe-PB18,NTHYen-CP19}. Nevertheless, it is not expected that this approach will give a complete description of the physical properties, in particular at the strong coupling limit.\par 
Motivated by this, we studied the ionic Hubbard model with mass imbalance in a one-dimensional lattice. In order to explore this system beyond mean-field theory, which predicts a weak first-order transition between a band insulator and an antiferromagnetic one, we used the  density matrix renormalization group (DMRG) algorithm. We found that the phase transition is  continuous and not a first-order one. Also, we observed that the phase transition is signaled by a maximum in the von Neumann block entropy, using this fact to build the phase diagrams of the model.\par 
This paper is organized as follows: In Sec.~\ref{sec2}, we give a brief description of the model considered in this investigation. The ground-state properties and the phase diagram determined by the maximum of block entropy for the ionic Hubbard with mass imbalance are shown in Sec.~\ref{sec3}. A summary of our results and main findings appears in the last section.
\section{\label{sec2} Model}
In this study, we consider two-color fermions with differing mass confined in a one-dimensional lattice with a two-site unit cell. A simple description of this system can be done through this  Hamiltonian:
\begin{align}
H=&-\sum_{i,\sigma}{t_{\sigma}\left(\hat{c}_{i,\sigma}^{\dagger}\hat{c}_{i+1,\sigma}+\hat{c}_{i+1,\sigma}^{\dagger}\hat{c}_{i,\sigma}\right)}\nonumber\\
& + \Delta\sum_{i,\sigma}{(-1)^{i}\hat{n}_{i,\sigma}}+U\sum_{i}{\hat{n}_{i\uparrow}\hat{n}_{i\downarrow}},
\label{H_imbalance}
\end{align}
\noindent where $\hat{c}_{i,\sigma}^{\dagger}$ and $\hat{c}_{i,\sigma}$ are the creation and annihilation operator of fermionic particles with spin $\sigma=\uparrow,\downarrow$ at site $i$ $(i = 1,\ldots,L)$. $\hat{n}_{i,\sigma}=\hat{c}_{i,\sigma}^{\dagger}\hat{c}_{i,\sigma}$ is the particle number operator for particles with spin $\sigma$ at site $i$, $t_{\sigma}$ is the hopping amplitude of a fermion with spin $\sigma$ to a near-nearest neighbor, $U$ is the local interaction between fermions with opposite spin, and $\Delta$ is the difference in on-site energies or staggered potential.\par 
\begin{figure}
\centering
\begin{tabular}{cc}\\
(a) &\\
 & \includegraphics[scale=0.4]{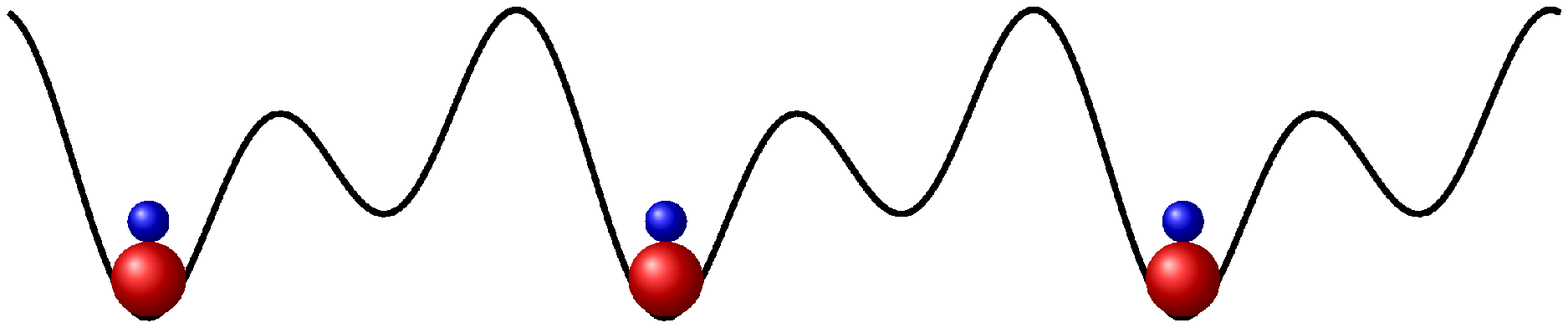}\\
(b) &\\
 & \includegraphics[scale=0.4]{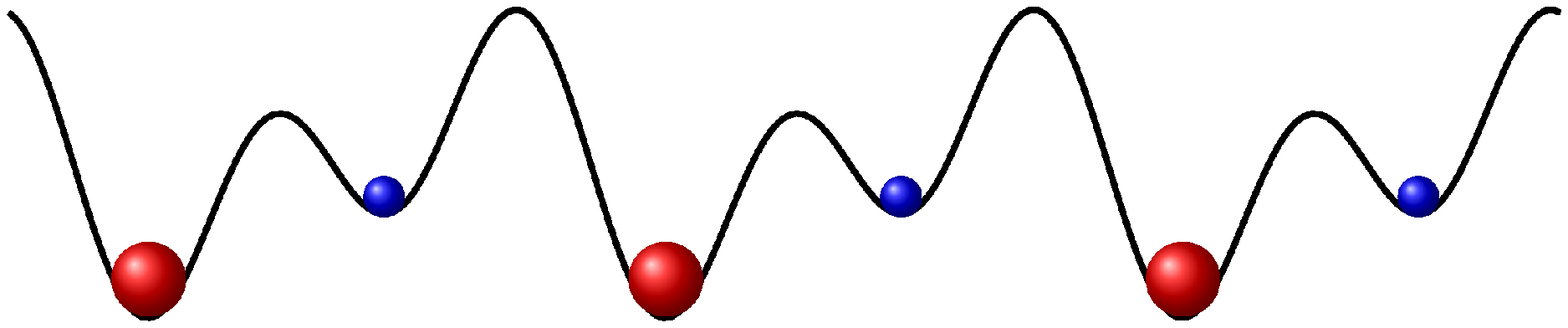}
\end{tabular}
\caption{Sketches of possible distributions of carriers in the ionic Hubbard model with mass imbalance. The red and blue spheres represent the particles with spin up and down, respectively. A band insulator state is displayed in (a), while a correlated insulator without mass imbalance appears in (b).}
\label{fig1}
\end{figure}
A mass-balanced situation implies that $t_{\uparrow}=t_{\downarrow}=t$, and if the lattice is homogeneous ($\Delta/t=0$), the Hamiltonian (\ref{H_imbalance}) corresponds to the Hubbard model that is the simplest model for describing electrons interacting in narrow bands~\cite{Hubbard-PRSL63}. This model is translational invariant and has spin $SU(2)$ symmetry; however if the staggered potential is different from zero or there is a mass imbalance in the system, these symmetries are explicitly broken,  giving rise to diverse phases in the system.\par
The well-known ionic Hubbard model is obtained when only the translational symmetry is broken ($t_{\uparrow}=t_{\downarrow}=t$, $\Delta/t>0$), which exhibits two phase transitions related to the competition between $U$ and $\Delta$~\cite{Manmana-PRB04}. For small values of the local interaction ($U/\Delta\ll 1$), a band insulator phase with a modulation of charge appears (see Fig.~\ref{fig1}-a), called a charge density wave (CDW). A Mott-insulator phase arises for a small value of the staggered potential, $U/\Delta\gg 1$ (see Fig.~\ref{fig1}-b). In the intermediate region between these phases, an engaging phase characterized by a dimerization between neighbor sites occurs.  This phase is called bond order (BO). The BO phase, together with the CDW-insulator, manifest the broken translational symmetry;  meanwhile, the Mott-insulator phase remains translational invariant in the charge sector.\par
If only the $SU(2)$ symmetry is explicitly broken ($t_{\uparrow}\neq t_{\downarrow}$, $\Delta/t_{\uparrow}=0$), an asymmetry in the spin sector appears. In fact, for the strong coupling limit, $U\gg t_{\uparrow},t_{\downarrow}$, the Hamiltonian (\ref{H_imbalance}) can be effectively mapped into the anisotropic $XXZ$ Heisenberg Hamiltonian~\cite{Grusha-IJMPB16}:
\begin{align}
H_{XXZ}&=J\sum_{i}\left({\hat{S}_{i}^{x}\hat{S}_{i+1}^{x}+\hat{S}_{i}^{y}\hat{S}_{i+1}^{y}+\gamma \hat{S}_{i}^{z}\hat{S}_{i+1}^{z}}\right),
\end{align}
\noindent where $\hat{S}^{x,y,z}$ are the spin-$\tfrac{1}{2}$ operators and  $J=4t_{\uparrow}t_{\downarrow}/U$, $\gamma=(t_{\uparrow}^{2}+t_{\downarrow}^2)/2t_{\uparrow}t_{\downarrow}$. Since $J>0$ and $\gamma>1$ for any value of $t_{\uparrow},t_{\downarrow}$, the system presents an antiferromagnetic ordering with a finite spin gap~\cite{DesCloizeaux-JMP66}.\par
When both symmetries, traslational and $SU(2)$, are broken, the system is fully described at the strong coupling limit by the following effective Hamiltonian~\cite{Grusha-IJMPB16,Gurzadyan2013}:
\begin{align}
H_{eff}&=\widetilde{J}\sum_{i}\left({\hat{S}_{i}^{x}\hat{S}_{i+1}^{x}+\hat{S}_{i}^{y}\hat{S}_{i+1}^{y}+\gamma \hat{S}_{i}^{z}\hat{S}_{i+1}^{z}}\right)\nonumber\\
&-h\sum_{i}{(-1)^i\hat{S}_{i}^{z}},
\label{H_eff}
\end{align}
\noindent where 
\begin{align}
\widetilde{J}=\frac{4t_{\uparrow}t_{\downarrow}U}{(U^2-\Delta^2)},\quad & h=\frac{2(t_{\uparrow}-t_{\downarrow})\Delta}{(U^{2}-\Delta^2)}. 
\end{align}
The last term in the Hamiltonian~(\ref{H_eff}), explicitly breaks the translational symmetry. The above effective Hamiltonian suggests that the ground state of the ionic Hubbard model with mass imbalance can exhibit modulation of charge with/without a finite spin gap.\par 
Motivated to study the ground state properties of the ionic Hubbard model under mass imbalance beyond mean-field theory, we numerically study the Hamiltonian (\ref{H_imbalance}), using the density matrix renormalization group (DMRG) method with the matrix product state (MPS) algorithm, in particular the ITensor library~\cite{itensor}. Considering a number of fermions $N_\sigma$ ($\sigma=\uparrow,\downarrow$) such that $L=N_{\uparrow}+N_{\downarrow}$ (half-filling), we determine the charge, spin,  and excitation gaps given by
\begin{align}
\Delta_{C}=&\: E_{0}(N_{\uparrow}+1,N_{\downarrow})+E_{0}(N_{\uparrow}-1,N_{\downarrow})\nonumber\\
&-2E_{0}(N_{\uparrow},N_{\downarrow})\\
\Delta_{S}=&\: E_{0}(N_{\uparrow}+1,N_{\downarrow}-1)-E_{0}(N_{\uparrow},N_{\downarrow}),\\
\Delta_{E}=&\: E_{1}(N_{\uparrow},N_{\downarrow})-E_{0}(N_{\uparrow},N_{\downarrow}),
\end{align}
\noindent $E_{0}(N_{\uparrow},N_{\downarrow})$ and $E_{1}(N_{\uparrow},N_{\downarrow})$ being the ground state and the first excited state energies, respectively.\\
\begin{figure}[t]
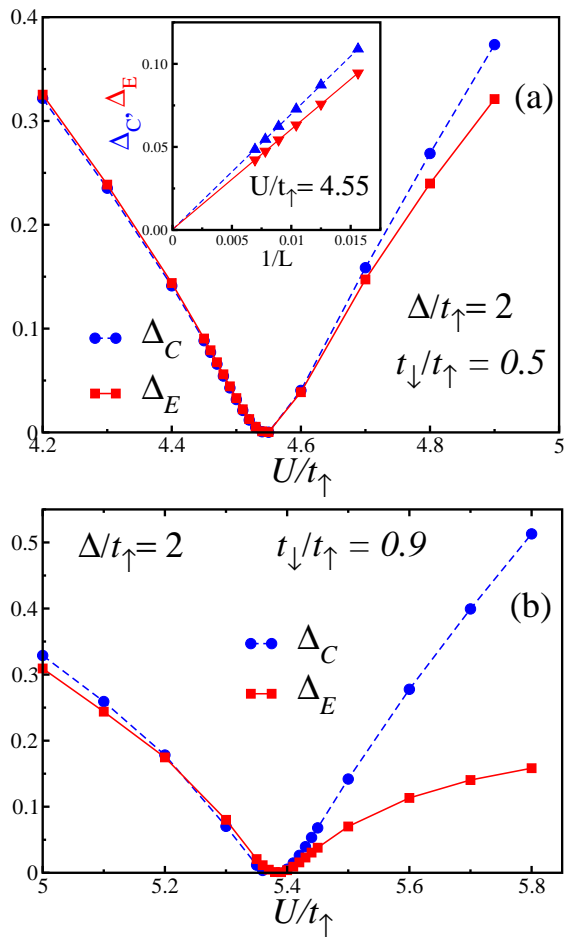

\centering
\begin{tabular}{c}
\includegraphics[scale=0.32]{Fig2a.eps}\\
\includegraphics[scale=0.32]{Fig2b.eps}
\end{tabular}
\caption{Extrapolated values of the charge ($\Delta_{C}$) and excitation ($\Delta_{E}$) gaps as a function of the local repulsion $U$ for an ionic Hubbard chain with band offset $\Delta/t_{\uparrow}=2.0$ and two different mass imbalance configurations: (a) $t_{\downarrow}/t_{\uparrow}=0.5$, and (b) $t_{\downarrow}/t_{\uparrow}=0.9$. The evolution of the charge and excitation gaps as the lattice size grows is displayed in the inset of Figure (a) for $U/t_{\uparrow}=4.55$. The lines in both figures are visual guides.}
\label{fig2}
\end{figure}
\section{\label{sec3} Ground State Properties}
In our numerical calculations, we set $t_\uparrow=1$ as our energy scale and consider lattices from $L=64$ to $L=512$ sites with open boundary conditions. In order to avoid meta-stable states, we use a noise parameter $a=10^{-5}\sim 10^{-12}$. To ensure a truncation error of around $10^{-9}$, we keep up to $800\sim 1000$ states per block. The error in the ground-state energy was $10^{-6}$ in the worst case, but around $10^{-8}$ or better in most cases.\par
\begin{figure}[t!]
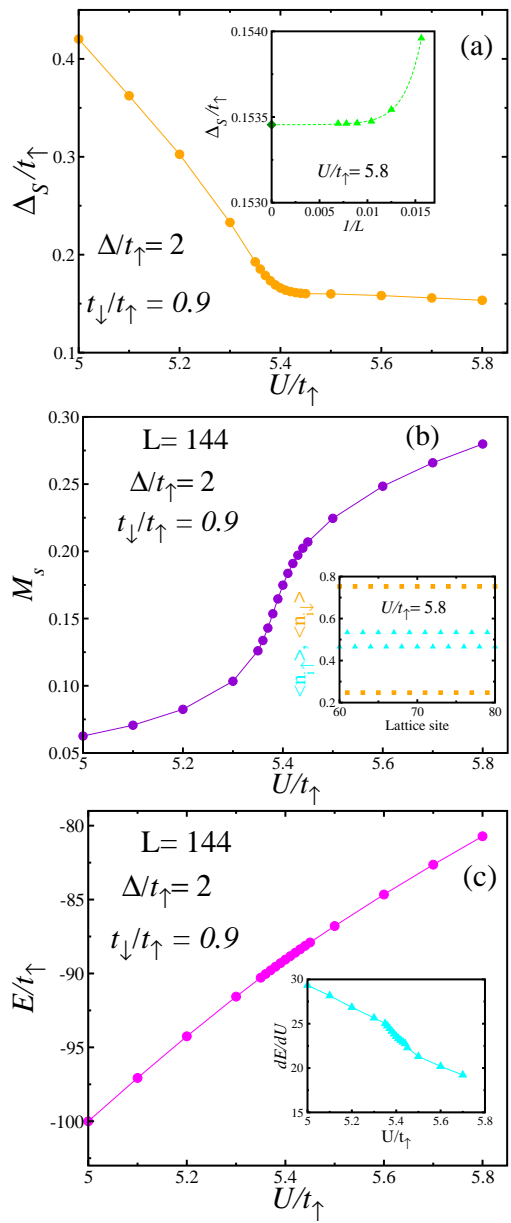

\begin{minipage}{18.9pc}
\includegraphics[width=15.7pc]{Fig3a.eps}
\end{minipage}
\hspace{5pc}%
\begin{minipage}{18.9pc}
\includegraphics[width=15.7pc]{Fig3b.eps}
\end{minipage}
\hspace{5pc}%
\begin{minipage}{18.9pc}
\includegraphics[width=15.7pc]{Fig3c.eps}
\caption{\label{fig3} Physical properties of an ionic Hubbard chain versus the local repulsion. Here the band offset and the mass imbalance are $\Delta/t_{\uparrow}=2.0$ and $t_{\downarrow}/t_{\uparrow}=0.9$, respectively. (Top) Extrapolated spin gap ($\Delta_{S}$) values. In the inset, we show the spin gap as a function of the inverse of the lattice size $1/L$. The extrapolated value at the thermodynamic limit is represented by a diamond. (Middle) The staggered magnetization for a lattice with $L=144$ sites. In the inset, we show the density profile for each spin orientation for $U/t_{\uparrow}=5.8$. (Bottom) Evolution of the ground-state energy for finite lattice. The second derivative of the ground-state energy appears in the inset. Except for the upper main plot, in all the others the points correspond to DMRG results. The lines are visual guides.}
\end{minipage}\hspace{2pc}%
\end{figure}
The evolution of the charge ($\Delta_C$) and excitation ($\Delta_E$) gaps as a function of the local interaction is shown in Fig.~\ref{fig2} for a chain with an energy offset $\Delta/t_{\uparrow}=2$, and two different values for the mass imbalance. In absence of a local interaction, it is expected that the charge and excitation gaps will be non-zero at the thermodynamic limit, due to the fact that the ground state corresponds to a charge density wave, whose unit cell is composed by one full site and another empty one. At this limit, both gaps coincide, and this fact remains when turning on the local interaction, but with a decrease in its value, as can be observed in Fig.~\ref{fig2}(a) ($t_{\downarrow}=0.5t_{\uparrow}$), following the scenario shown before in the balanced case~\cite{Manmana-PRB04}. In the inset of Fig.~\ref{fig2}(a), we show the charge and excitation gaps as a function of the inverse of the lattice size, considering $t_{\downarrow}=0.5t_{\uparrow}$ and $U^{*}/t_{\uparrow}=4.55$. We observe that both gaps decrease linearly reaching a tiny number treated as zero at the thermodynamic limit, indicating that at this point a quantum phase transition can be happening. So, the evolution of the extrapolated values of $\Delta_C$ and $\Delta_E$ as a function of $U$ changes at  $U/t_{\uparrow}=4.55$, from which both gaps increase monotonously, separating as the interaction grows (Fig.~\ref{fig2}(a)) and indicating a different ground state. In Fig.~\ref{fig2}(b), we consider $t_{\downarrow}=0.9t_{\uparrow}$ and display the evolution of the extrapolated values of $\Delta_C$ and $\Delta_E$ versus the local interaction, observing a  similar global scenario  with both gaps decreasing (increasing) before (after) a certain value of the local interaction, where both vanish. In the latter case, we obtained that $U^{*}/t_{\uparrow}=5.38$, indicating that the critical points move to larger values as $t_{\downarrow}/t_{\uparrow}$ increases. Comparing the above critical local parameters with ones forecast by the mean-field calculations $U^{*}_{MF}/t_{\uparrow}=4.25$ and $4.81$ for $t_{\downarrow}/t_{\uparrow}=0.5$ and $0.9$, respectively, it is clear that correlations move the critical points for larger values, a displacement that depends on the mass asymmetry.\par 
In order to acquire more information about our system, we display other physical quantities in Fig.~\ref{fig3}, taking into account an energy offset of $\Delta/t_{\uparrow}=2$ and $t_{\downarrow}=0.9t_{\uparrow}$. The behavior of the spin gap as a function of $1/L$ appears in the inset of Fig.~\ref{fig3}(a) for $U/t_{\uparrow}=5.8$, and it is clear that the spin gap decreases as the lattice size grows, such that at the thermodynamic limit, the spin gap reaches a non-zero finite value $\Delta_S/t_{\uparrow}=0.153$ (dark green diamond point). Repeating this procedure, we obtained the main plot of Fig.~\ref{fig3}(a), in which the extrapolated values of the spin gap decrease monotonously as the local interaction increases up to the critical point $U^{*}/t_{\uparrow}=5.38$ (charge and excitation gap vanishes here), where the spin gap changes drastically, reducing its rate of change and tending to a fixed value for large values of $U$. The main result unveiled here is that both phases have a non-zero spin gap, being small for the phase that is above the critical point. Note that in the balanced case a phase with zero spin gap is established after the critical point.\par
\begin{figure}[t] 
\includegraphics[width=18pc]{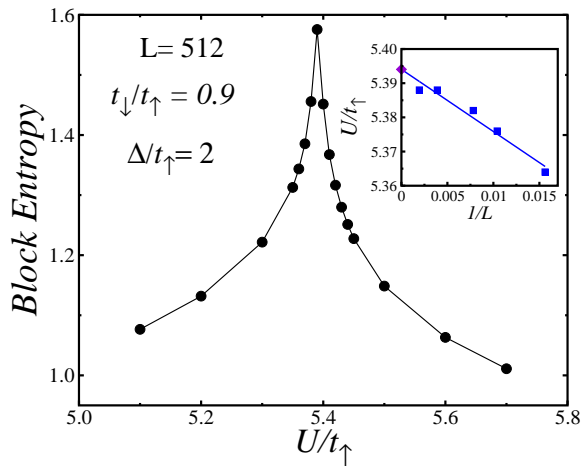} 
\caption{\label{fig4} Von Neumann block entropy as a function of the local repulsion $U$ for an ionic Hubbard chain with $L=512$ sites, a band offset $\Delta/t_{\uparrow}=2.0$, and mass imbalance  $t_{\downarrow}/t_{\uparrow}=0.9$. The evolution of the maximum value of the von Neumann block entropy as the lattice size grows is displayed in the inset of figure. The lines are visual guides.}
\end{figure} 
Another interesting quantity to follow as $U$ grows is the staggered magnetization given by  
\begin{align}
M_{s}&=\frac{1}{L}\sum_{i}{(-1)^i\Bigl(\langle \hat{n}_{i,\uparrow} \rangle - \langle \hat{n}_{i,\downarrow}\rangle\Bigr)},
\end{align}
\noindent which appears Fig.~\ref{fig3}(b) for a lattice with a size of $L=144$ sites. From zero and small values of the local interaction, the ground state is a CDW with a two-site unit cell, one of them full and the other empty. For this ground state, the magnetization is zero; however, the magnetization grows slowly as we approach the critical point, changing its behavior at this point, confirming that a phase transition takes place. After the critical point, the staggered magnetization grows monotonously with the local interaction, which indicates a redistribution of charge in the unit cell, and an antiferromagnetic order is established. Hence after the critical point the ground state is antiferromagnetic with a finite spin gap, a scenario that has also arisen in other systems~\cite{Hikihara-PRB03,Dagotto-PRB92}. In the inset of Fig.~\ref{fig3}(b), we display the density profile for each spin orientation for a local repulsion $U/t_{\uparrow}=5.8$, keeping the other parameters as in the main figure. We are inside of the antiferromagnetic region, and observe that both $\langle \hat{n}_{i,\uparrow} \rangle$ and $\langle \hat{n}_{i,\downarrow}\rangle$ oscillate out of phase with each other throughout the lattice, establishing a modulation of charge with a unit cell with two sites. Although at each site the total occupation is lower or higher than one, at the unit cell we have exactly two particles.\par
Our previous numerical calculations and the mean-field study of the ionic Hubbard model with mass imbalance indicate that this model exhibits only two phases, in contrast with the three phases found in the balanced-model. A central question about a phase transition is its order, which the mean-field calculations suggest is first order. To contribute to this subject, we show the evolution of the ground-state energy as a function of the local interaction for a lattice with $\Delta/t_{\uparrow}=2$, $t_{\downarrow}=0.9t_{\uparrow}$, and $L=144$ (see Fig.~\ref{fig3}(c)). The ground-state energy increases monotonously with $U$ and no discontinuity is observed, which indicates that this transition is not first-order, contradicting the mean-field findings. In the inset of Fig.~\ref{fig3}(c), the derivative of the ground-state energy appears, from which it is clear that it  is continuous around the critical point, leading to the conclusion that the quantum phase transition in the ionic Hubbard model with mass imbalance is continuous.\par 

Nowadays, it is well known that the tools of quantum information theory are useful for identifying critical points without a \textit{piori} knowledge about it~\cite{Amico-RMP08}. The position of the critical points emerges by means of characteristic behaviors of the quantum information witnesses around it, for instance extreme values or singularities. To identify the critical points of the ionic Hubbard model with mass imbalance, we will use the von Neumann block entropy, which is cheaper than energy gaps.\par
\begin{figure}[t!]
\begin{minipage}{18.pc}
\includegraphics[width=17.0pc]{Fig5a.eps}
\end{minipage}
\hspace{5pc}%
\begin{minipage}{18.pc}
\includegraphics[width=17.0pc]{Fig5b.eps}
\end{minipage}
\hspace{5pc}%
\begin{minipage}{18.pc}
\includegraphics[scale=0.55]{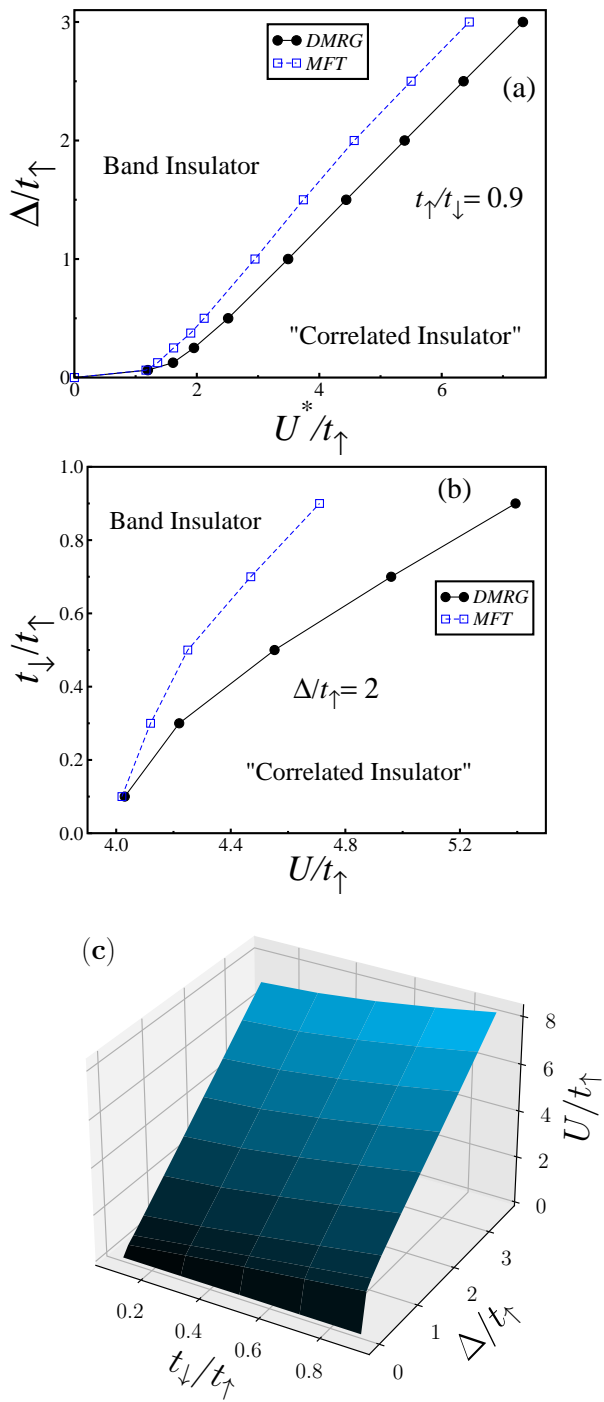}
\caption{\label{fig5} Phase diagrams of the ionic Hubbard chain with mass-imbalance. 
(Top) Band offset versus local repulsion plane with $t_{\downarrow}/t_{\uparrow}=0.9$. (Middle) $t_{\downarrow}-U$ plane for a band offset $\Delta/t_{\uparrow}=2$. In the above figures, the squares are critical points obtained by mean-field calculations~\cite{Sekania-PRB17}, whereas the circles correspond to extrapolations from DMRG results. (Bottom) Local critical repulsion as a function of the band offset and the mass-imbalance for a Hubbard chain at half-filling. The lines are visual guides.}
\end{minipage}\hspace{2pc}%
\end{figure}
We consider a system with $L$ sites divided into two parts. Part $A$ has $l$ sites ($l=1,...,L$), and the rest form part $B$, with $L-l$ sites. The von Neumann block entropy of block $A$ is defined by $ S_A=-Tr\varrho_A  ln  \varrho_A$, where $\varrho_A=Tr_B \varrho$ is the reduced density matrix of block $A$ and $\varrho= |\Psi\rangle \langle \Psi|$ the pure-state density matrix of the whole system. Namely, we consider $l=L/2$, which is the von Neumann block entropy. In Fig.~\ref{fig4}, we show the block entropy as a function of the local interaction for a chain with the parameters $\Delta/t_{\uparrow}=2$, $t_{\downarrow}=0.9t_{\uparrow}$ and $L=512$. As expected, the block entropy grows monotonously as $U$ increases, due to the fact that without interaction, the ground state is a CDW, which is a separable state leading to a value of zero for the block entropy. Turning on the local interaction, the charge fluctuations increase, as well as the entanglement, as can be seen in the block entropy, which reaches a maximum value at $U^{*}/t_{\uparrow}\sim5.38$, matching the position for which other quantities exhibit an anomalous behavior. After this maximum, the block entropy decreases, due to the fall in the charge fluctuations in the correlated insulator phase, for which we expected a finite value of the block entropy for larger values of local repulsion. Therefore, we observe that the von Neumann block entropy gives us the critical point position where the quantum phase transition takes place. In the inset of Fig.~\ref{fig4}, we display the maximum position of the block entropy as the lattice size increases, observing a small displacement to lower values, which leads to the critical point $U^{*}/t_{\uparrow}=5.39$ at the thermodymanic limit (violet diamond point).\par  
Replicating the above procedure for other values of the energy offset and keeping fixed $t_{\downarrow}=0.9t_{\uparrow}$, we obtain the phase diagram shown in Fig.~\ref{fig5} (a), where the critical points found by mean-field and DMRG are shown in the plane $\Delta-U^{*}$. Starting from $\Delta=0$, for which $U^{*}$ vanishes, we observe that the critical points increase quickly for small values of $\Delta$ and then follow an almost linear relation for both mean-field and DMRG results. It is clear from the figure that when considering the correlations in a more precise way (DMRG results), the critical points move towards higher values, a tendency that increases when the energy offset grows. The line critical points separates the band insulator phase (left region) from the correlated insulator phase (right region). Fixing the energy offset ($\Delta/t_{\uparrow}=2$), we observe the monotonous growth of the critical points for both the mean-field and the DMRG results (see Fig.~\ref{fig5} (b)), which only coincide for extreme mass imbalance. The complete phase diagram for the ionic Hubbard model with mass imbalance is shown in Fig.~\ref{fig5} (c), where the surface of the critical points  divides the space, leaving a band insulator (correlated insulator) phase above (underneath) it.\par
\section{\label{sec4} Summary}
Motivated by the mean-field findings about the ionic Hubbard model with mass imbalance, we reviewed this model at half-filling, using the numerical technique density matrix renormalization group. Several quantities related to the ground state were calculated, such as the energy gaps (charge, spin and excitation), the staggered magnetization, and the von Neumann block entropy. Looking over the space parameters, we were able to identify two insulator phases, both with a two-site unit cell: a band insulator and a correlated  insulator with antiferromagnetic order. The latter phase has a finite non-zero spin gap generated by the asymmetries of the model, this being a curious result, which has also been found in other systems~\cite{Hikihara-PRB03,Dagotto-PRB92}.\par 
To characterize the quantum phase transition from the band insulator to the correlated one, we follow the evolution of the ground-state energy throughout this transition, finding that it is continuous, contradicting the insight obtained with the mean-field calculations. The critical points of the band-correlated transition were determined with the von Neumann block entropy, which exhibits a maximum around the separation point.  After extrapolating to the thermodynamic limit, we observed that our critical points are larger than the mean-field ones, except for very small band offsets and extreme mass imbalance.\par 
\section*{Acknowledgments}
This research was supported by DIEB- Universidad Nacional de Colombia (Grant No. 51116).
 Author Contribution Statement:
D. C. Padilla-Gonz\'alez programmed the DMRG code and carried out the calculations. R. Franco contributed to the discussions. J. Silva-Valencia planned and designed the study. All contributed to writing the paper.

\bibliography{/home/jereson/PAPERS/Bib/Bibliografia.bib}

\end{document}